\documentclass[preprint,12pt]{elsarticle}

\usepackage{amssymb}
\def \i{\mathrm{i}}
\def \e{\mathrm{e}}

\journal{Physica C}

\begin{document}

\begin{frontmatter}


\ead{jorge.berger@braude.ac.il}

\title{Attempt to describe phase slips by means of an adiabatic approximation}


\author[inst1]{Jorge Berger}
\author[inst2]{Edson Sardella}

\affiliation[inst1]{organization={Department of Physics and Optical Engineering, Braude College},
            city={Karmiel},
            postcode={2161002}, 
            country={Israel}}

\affiliation[inst2]{Departamento de Fisica, Faculdade de Ciencias, Universidade Estadual Paulista (UNESP),Bauru-SP,17033-360,Brazil}

\begin{abstract}
In the description of non equilibrium situations in a superconductor at temperatures far below its critical temperature, the Keldysh--Usadel technique (KUT) is required. However, the non-stationary KUT has not been applied to realistic circuits. Moreover, the stationary KUT has been applied in situations where time independence is not guaranteed. As a plausibility check for this procedure, we resort to a toy model (the Ginzburg--Landau model), in situations that involve very slow evolution. We find that, even in these situations, neglecting the explicit influence of time variation leads to inaccurate or even qualitatively wrong description of phase slips. 
\end{abstract}


\begin{highlights}
\item Our results indicate that, in the analysis of superconducting circuits in which phase slips could be present, the full non-stationary formalism has to be used.
\end{highlights}


\end{frontmatter}


\section{Introduction}
\label{Intro}
Superconducting nanocircuits and nanodevices have been available for decades \cite{Fo,We,optics,Carmine,LevS,KoeRev}. Here we focus on their most basic component: a nanowire. When a voltage is applied between the ends of a superconducting wire, a phase difference between them builds up. Eventually, a phase slip (or a set of phase slips) occurs, i.e., superconductivity vanishes somewhere in the wire, enabling discontinuities in the phase by some multiple of $2\pi$ \cite{PS}. 

Phase slips (PSs) manifest themselves as increments of the wire resistance, but not only. An individual PS center can be controlled by radiation \cite{science} and can be sensed by means of tunneling of superconducting pairs \cite{Jackel}.  PSs could perform as qbits \cite{Q}. PSs may lead to false counts in single-photon detectors \cite{IEEE} and exert back action on measuring instruments \cite{stat}.  From the general point of view of dynamic systems and phase transitions, there is special interest in systems that can evolve either to a stationary regime, without PSs, or to a periodic regime, with PSs, depending on some control parameter \cite{Koby}. It should be noted that PSs are not imperative for non-zero resistance \cite{lossless} or for discontinuities in the voltage-current characteristic of a superconducting wire \cite{Koby}. In this study we do not consider PSs that are due to thermal \cite{thermal} or quantum \cite{q1,q2} fluctuations.

Since PSs are a time dependent phenomenon, their analysis requires a theory that takes time dependence into account. An available theory is the Kramer--Watts-Tobin model  (KWT)\cite{KW1,KW2,Ivlev}. Among the numerous studies based on KWT, some consider infinite wires \cite{Rangel}, some consider finite wires carrying a fixed current \cite{Koby,me}, and some consider finite wires that withstand a fixed voltage \cite{physc}. It should be noticed that the voltage-current characteristic of a superconducting wire in the case that the current is kept constant differs qualitatively from the case in which the voltage is kept constant \cite{Micho1,Micho2}. 
Unfortunately, the KWT model is valid only close to the critical temperature $T_c$, whereas the most interesting temperature range for application purposes is far below $T_c$. In this range, the available theory is the quasiclassical Keldysh--Usadel technique (KUT) \cite{Belzig,Kamenev}, that relies on a one-time Green’s function formalism. KUT has been used to describe several systems and effects, such as short and long SNS junctions \cite{Belzig}, interference effects in normal metals \cite{Belzig}, tunneling barriers \cite{Belzig}, superconducting quantum interference proximity transistors \cite{SQUIPT}, multiterminal structures \cite{Brink}, and various heterostructures that may include superconducting and ferromagnetic components \cite{cph,Buzdin}. The drawback of KUT is that the applicability of the one-time Green’s function formalism is limited to stationary situations; time dependence would involve two-time Green’s functions, that deter researchers from this formalism. Nevertheless, stationary KUT is frequently used in situations in which time independence is not guaranteed \cite{Boo,Keizer,Pekola,Andreev}, but is rather assumed a priori. The use of the stationary formalism is sometimes justified with arguments such that iterations converge, or the fact that measurements show a stationary behavior; however, measurements cannot distinguish between a time independent voltage and a periodic voltage with a period that is much shorter that the voltmeter response time.

We are therefore left with two options: a theory that is not valid in the relevant temperature range, and a theory that is not applicable to the relevant phenomenon. An approach that is called for is the extension of KUT to non-stationary situations by means of an adiabatic approximation (AA) \cite{Golub,tunnel}, namely, to consider situations in which the state of the system evolves sufficiently slowly to enable a reasonable description by means of a theory that ignores time dependence. The present study provides a test to this approach.

In order to evaluate how close the results of stationary KUT are to those of a valid time dependent theory for a given situation, the time dependent theory is required. Since at present we are unable to use the required theory, we will resort to the range in which a time-dependent theory is available (i.e. KWT). Since the behavior of a superconducting wire is similar whether slightly below or far below $T_c$, the success or failure of the AA in one of these ranges provides an educated guess for the expected suitability of this approximation in the other range.

The system we consider consists of a thin superconducting wire that connects two conducting banks. The banks are at equilibrium and kept at fixed potentials. We will consider two candidates for slow evolution. The first case will be that of small applied voltage, so that the Josephson frequency is small. The second case will be that of continuous passage between a stationary and an oscillatory regime: for parameters near the stationary regime, we expect quasi-stationary behavior, i.e. the oscillation period diverges.

\section{Mathematical model\label{model}}
For a one-dimensional system, the vector potential can be taken as zero. Then the evolution equation becomes \cite{KW1}
\begin{equation}
    \frac{u}{[1+\gamma^2|\psi |^2]^{1/2}}
    \left[\frac{\partial}{\partial t}+\i\mu+\frac{\gamma^2}{2}\frac{\partial |\psi |^2}{\partial t}\right]\psi = \frac{\partial^2 \psi}{\partial x^2}+(1-|\psi |^2)\psi \,.
    \label{KWevol}
\end{equation}
Here $\psi$ is the order parameter, that determines the supercurrent density, $u$ and $\gamma$ are material parameters, $x$ is the position in units of the coherence length $\xi$, $t$ is the time in units of $\xi^2/Du$, where $D$ is the diffusion coefficient, and $\mu$ is the electrochemical potential in units of $\hbar Du/2e\xi^2$. The current density $j$, in units of $\hbar\sigma Du/2e\xi^3$, where $\sigma$ is the normal conductivity, is given by
\begin{equation}
    j={\rm Im}[\psi^*\,\partial\psi/\partial x] -\partial\mu/\partial x\,,
    \label{j}
\end{equation}
and invoking electroneutrality,
\begin{equation}
   \partial^2\mu/\partial x^2=\partial{\rm Im}[\psi^*\,\partial\psi/\partial x]/\partial x\,.
   \label{neutral}
\end{equation}

Since we are not interested in the accuracy of the model, but rather in how close the AA reproduces it, and since varying $\gamma$ does not lead to qualitative changes in the regime diagram \cite{me}, we set $\gamma =0$ and Eq. (\ref{KWevol}) becomes the time-dependent Ginzburg--Landau equation. Moreover, from the discussion in \cite{physc} we can conclude that the influence of $u$ is also weak and take $u=1$. Thus, Eq. (\ref{KWevol}) reduces to
\begin{equation}
        \left[\frac{\partial}{\partial t}+\i\mu\right]\psi = \frac{\partial^2 \psi}{\partial x^2}+(1-|\psi |^2)\psi \,.
    \label{TDGL}
\end{equation}
This is the evolution equation used in \cite{physc}. 

Equations (\ref{neutral}) and (\ref{TDGL}) have to be complemented with boundary conditions. For a wire of length $2L$ with applied voltage $V$, we write $-L\le x\le L$ and take the potential equal to $\pm V/2$ at the ends. Consistently, the order parameter at the banks has to be $r\exp (\mp\i Vt/2)$, where $r$ represents the superconductivity strength of the banks; in particular, if the banks are made of the same material as the wire, then $r=1$, and if they are normal metals, then $r=0$. In summary, we take the boundary conditions
\begin{equation}
    \mu (t,\pm L)=\mp V/2\,, \;\;\; \psi (t,\pm L)=r\e ^{\pm \i Vt/2}\,.
    \label{bc}
\end{equation}

Equations (\ref{neutral})--(\ref{bc}) constitute our time-dependent model.
The following sections investigate whether there are situations in which $\partial\psi /\partial t$ can be neglected, so that Eq.\ (\ref{TDGL}) can be replaced by the stationary equation
\begin{equation}
        \i\mu\psi = \frac{\partial^2 \psi}{\partial x^2}+(1-|\psi |^2)\psi \,.
    \label{STDGL}
\end{equation}
Replacement of (\ref{TDGL}) with (\ref{STDGL}) amounts to an adiabatic approximation. In the AA, time enters the problem through the boundary conditions only.

 When comparing solutions obtained using (\ref{STDGL}) with those obtained using (\ref{TDGL}), it has to be borne in mind that the transformation $\{t\rightarrow t+2\pi/V,\psi\rightarrow -\psi\}$ leads to an equivalent solution.

We can think of two cases in which $\psi$ changes slowly. The obvious case is that of a small applied voltage between the ends of the wire, leading to a low Josephson frequency. A more interesting situation is that of a bifurcation, i.e.\ there is a critical value of $V$ or $L$, such that below this value there is a stationary regime, whereas above this value there is a periodic regime; continuity then requires very long periods in time closely above the critical value.
In the following examples, we will take $L=4$ (which corresponds to a wire of length of about a third of a micrometer for Nb at 0K).

\section{Superconducting banks}

This case is implemented by taking $r=1$ in the boundary conditions (\ref{bc}). We know in advance that PSs must occur in order to avoid divergence of the gradient of the phase in the wire. From the Josephson relation we foresee a periodic regime that in our units has a period of time $2\pi/V$, and, in order to have a slowly varying state, we require a small value of $V$. We take $V=0.01$.

Our numerical procedure is described in \ref{appnum}.

The results that we present for the time-dependent model [i.e.\ using (\ref{TDGL})] correspond to times after convergence to a clearly defined regime. Equations (\ref{neutral})--(\ref{bc}) support the symmetry $\psi (t,-x)=\psi^*(t,x)$, $\mu (t,-x)=-\mu (t,x)$, and the results we obtained do not break this symmetry, so that it suffices to present our results for $x\ge 0$. 

\begin{figure}
\centering
\includegraphics[width=0.93\linewidth]{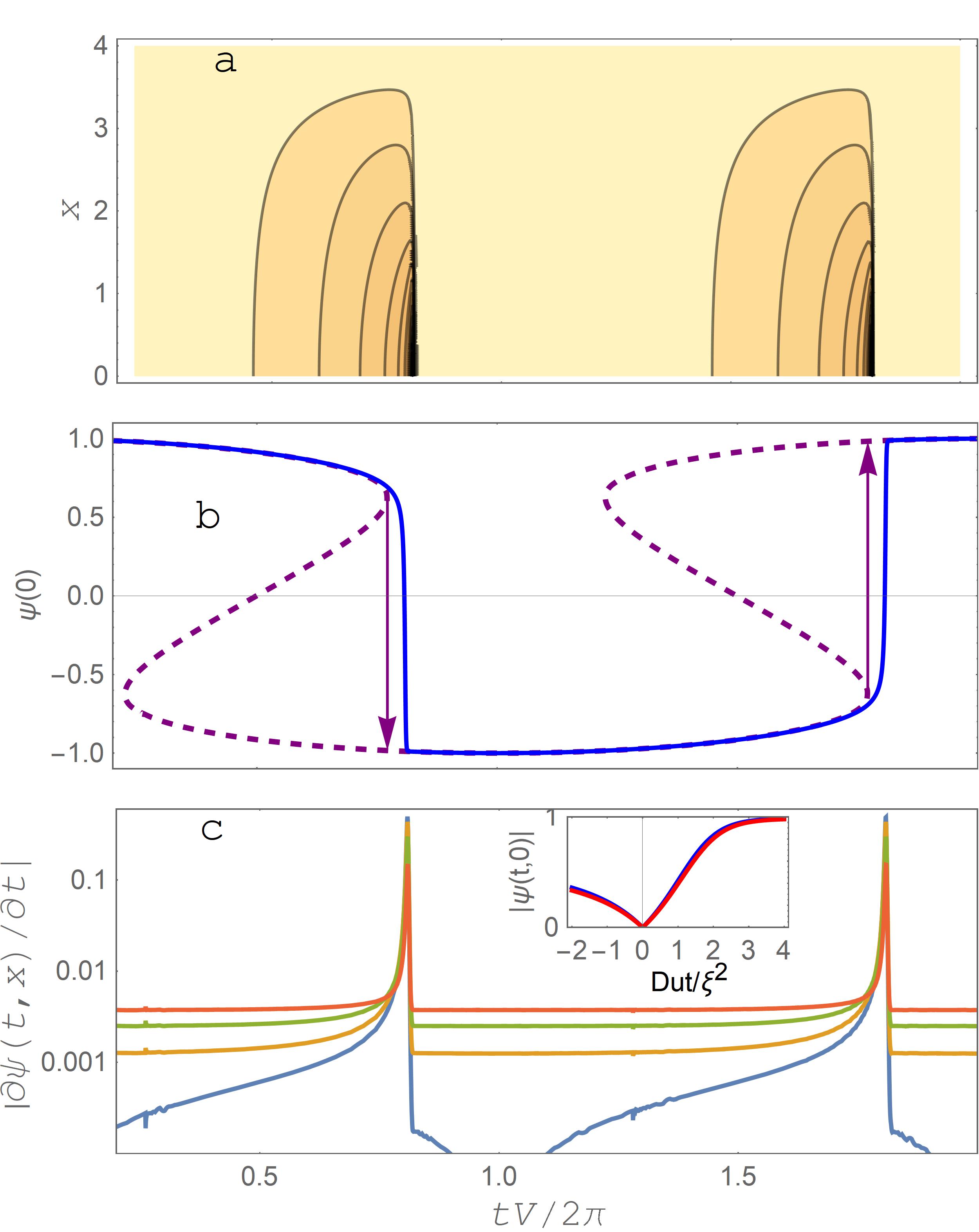}
\caption{Case $r=1$, $V=0.01$. The abscissa is common to all panels. \textbf{a} Contour plot of $|\psi (t,x)|$. Darker colors stand for smaller $|\psi |$. \textbf{b} $\psi (t,0)$ according to the time-dependent model (continuous, blue) and to the adiabatic approximation (dashed, purple). The arrows indicate discontinuous passage to another branch. \textbf{c} $|\partial\psi /\partial t|$ as a function of $t$. Blue: $x=0$; orange: $x=1$; green: $x=2$; red: $x=3$. Inset: $|\psi (t,0)|$ close to the phase slip for different voltages. The origin of the time is taken at the phase slip. Note that time is not scaled by $V$ in the inset. Blue: $V=10^{-2}$; red: $V=10^{-3}$.}
\label{fig:smallV}
\end{figure}

Figure \ref{fig:smallV}a is a contour plot of $|\psi (t,x)|$ (according to the time-dependent model). As expected from \cite{physc}, there are PSs at the center of the wire, that occur periodically with a period $2\pi /V$. Figure \ref{fig:smallV}b compares $\psi (t,0)$ according to the time-dependent model (continuous line) with $\psi (t,0)$ according to the AA (dashed line). Note that $\psi (t,0)$ is real. Since at the boundaries $x=\pm L$ there is no difference between the model and the approximation, the agreement is expected to be better for $x\neq 0$. We see that there are time ranges where the AA yields three solutions, so that in order to compare with the model we require some interpretation. The natural interpretation regards the middle branch as a set of unstable solutions, whereas solutions in the upmost and the lowermost branches can be stable. At times for which an end of a stable branch is reached, the state of the system has to jump to the other branch, as shown by the purple arrows. We see that the approximated and the model solutions are practically indistinguishable, except close to the PSs, where the decay of the model solution is gradual. We may now inquire for the origin of the discrepancy introduced when we drop $\partial\psi /\partial t$, which is expected to be small for $V\ll 1$. The answer is provided by Fig.\ \ref{fig:smallV}c: although $\partial\psi /\partial t$ is small most of the time, it is not small close to the PSs.

Figure \ref{fig:smallV}c tells us that PSs remain a fast process, even for $V\ll 1$. This result is expected from (\ref{TDGL}), because $V$ enters the equation only through the term $\i\mu\psi$. Near a PS, $\psi$ is small and, when also $\mu$ is small, this term becomes negligible and the applied voltage has no influence on the evolution of $\psi$. Indeed, the inset in the figure indicates that, for $V\lesssim 10^{-2}$, $\psi$ is practically independent of $V$ near a PS.

The time-average of the current density is roughly 10\% smaller in the case of the AA than for the time-dependent model. 

\begin{figure}
\centering
\includegraphics[width=0.9\linewidth]{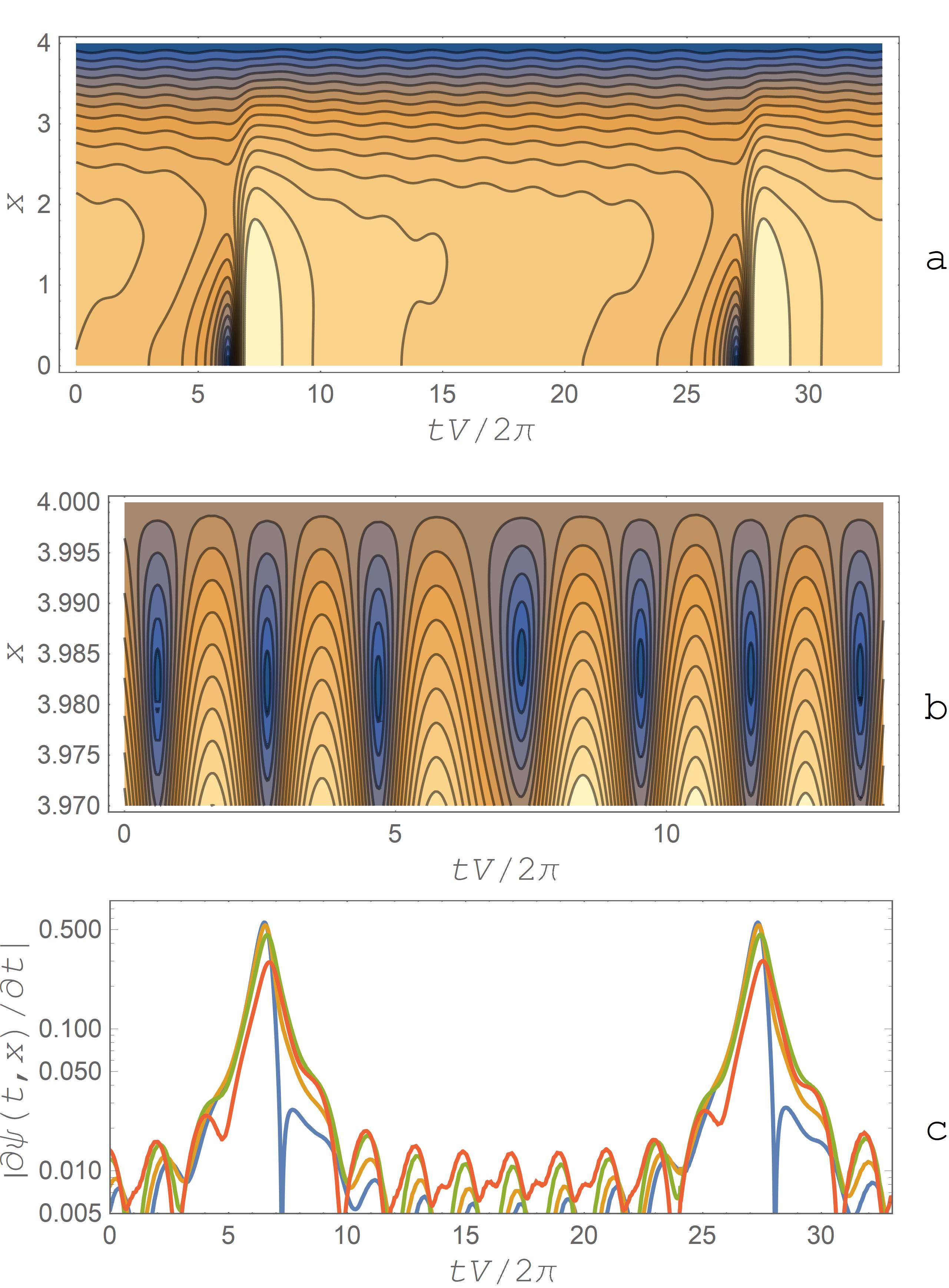}
\caption{Case $r=0.01$, $V=2.22$. \textbf{a} Contour plot of $|\psi (t,x)|$ that contains two significant phase slips at $x=0$. \textbf{b} Examples of irrelevant phase slips, that quasi-periodically occur at a distance of $\sim 0.02$ from each of the boundaries, with a repetition time of $\sim 4\pi/V$. The maximum value of $|\psi |$ in this panel is about 30 times smaller than the maximum in panel \textbf{a}. Panel \textbf{b} is actually part of panel \textbf{a}, but is not visible in the latter.  \textbf{c} $|\partial\psi /\partial t|$, with the same color meanings as in Fig.\ \ref{fig:smallV}c.}
\label{fig:222}
\end{figure}

\section{Quasi-normal banks}
A normal metal bank is usually analyzed by taking $r=0$. However, in this case the phase at the boundaries becomes undefined, so that the information about the Josephson evolution of the phase is lost. In order to have well defined phases, we will consider instead quasi-normal banks, such that $0<r\ll 1$. Quasi-normal boundary conditions are not less realistic than strictly normal boundary conditions, since the assumption that the points at $x=\pm L$ are in equilibrium despite the fact the current density does not vanish at them is an idealization anyway. In the following, we take $r=0.01$.

When dealing with quasi-normal banks, we have to distinguish between significant PSs, that take place where $|\psi (t,x)|$ is usually large, and irrelevant PSs, at which $\psi (x,t)=0$ for values of $x$ such that the order parameter is anyway small at all times. Examples of significant PSs are shown in Fig.\ \ref{fig:222}a and examples of irrelevant PSs are shown in Fig.\ \ref{fig:222}b. 

By solving numerically Eqs.\ (\ref{neutral})--(\ref{bc}) in the range $0\le V\le 3.55$, we find that there is a critical voltage $V_c\approx 2.195$ such that for $V<V_c$ there are no significant PSs, whereas for $V>V_c$ a PS occurs at $x=0$, with repetition time 
\begin{equation}
    T\approx \beta/\sqrt{V-V_c}\,,
\label{T}
\end{equation}
where $\beta =9.22$. It follows that the most promising region for success of the AA lies above and close to $V_c$. 

\subsection{Range $V<V_c$}
We performed calculations for $V=2.15$, for a long period of time. At the boundaries, $|\partial\psi /\partial t|=rV/2$, and we found that $|\partial\psi /\partial t|$ decreases as $|x|$ decreases. Accordingly, $\partial\psi /\partial t$ is small everywhere, and we can expect that the AA should lead to accurate results.

We found that, for fixed $x$, $\psi (t,x)$ is a function of time that slightly oscillates around an average value. The time-average values of $|\psi|$ for the time-dependent model and for the AA coincide within 0.01 units, but the amplitude of the oscillations can be up to ten times larger in the case of the approximation.

\subsection{Range $V>V_c$}
Figure \ref{fig:222} presents some of our results for $V=2.22$, slightly above $V_c$. The repetition time is $\sim 59$, naively suggesting a very slow variation. Unfortunately, Fig.\ \ref{fig:222}c shows that $\partial\psi /\partial  t$ is not small during a large fraction of the time, much larger than in the case of Fig.\ \ref{fig:smallV}c.

Before we go on, let us spell out what we could expect. Let us assume that significant PSs occur only at $x=0$. In the AA, time enters the problem only through Eq.\ (\ref{bc}), which just covers a time period of extent $4\pi /V$. If the time lapse between consecutive PSs is $4n\pi /V$, where $n$ is a natural number, then there must be at least $n$ branches of stable solutions with $\psi (t,0)>0$, such that the passage between branches is determined by continuity, and $n$ branches with $\psi (t,0)<0$, so that there are at least $2n$ stable branches. For a time lapse in the range between $4(n-1)\pi /V$ and $4n\pi /V$, there should be at least $2n$ branches, where one of them becomes unstable before filling the entire range allowed by the periodicity of the boundary conditions. In a polar graph where $|\psi (t,0)|$ is the radius and $Vt/2$ is the angle, a PS would be described as part of an orbit that spirals into the origin, possibly backwards in time.

According to the previous paragraph, the AA would require about 20 branches of solutions in order to describe, at least qualitatively, the results in Fig.\ \ref{fig:222}a. However, we found just three branches for $V=2.22$. For one of them $\psi (t,0)\sim 0.7$, there is a twin branch for $\psi (t,0)<0$, and there is a branch such that the wire is quasi-normal everywhere [$\psi (t,x)$ is of the order of $10^{-2}$ and irrelevant PSs occur with period $2\pi/V$]. 

We looked for solutions such that $0.05\le |\psi (t,0)|\le 0.65$ for some $t$, and none was found. Within numeric accuracy, the orbits for $|\psi (t,0)|\sim 0.7$ are closed. These results indicate that if there exists a critical voltage $V_c$ in the AA, then $V_c>2.22$.

In \ref{timebetween} we tried to find the value that the critical voltage would have in the AA. For this purpose, we did not drop the time derivative in (\ref{TDGL}), but rather multiplied it by a factor $\alpha$, with the intention of obtaining the critical voltage as a function of $\alpha$, and the AA result as a limit. We found, however, that the critical voltage is independent of $\alpha$. What does depend on $\alpha$ is the value of $\beta$ in (\ref{T}), and in the AA limit appears to become unphysical. In summary, the AA is unable to describe the presence of a critical voltage.

\section{Discussion}
Under the assumption that the Kramer--Watts-Tobin model provides a qualitative representation of how a superconducting state evolves in a nanowire,
an adiabatic approximation can fairly describe the state of a nanowire for low applied voltages,  away from the moments at which PSs occur. This approximation could also describe the superconducting state in the case of quasi-normal banks, as long as there are no PSs far from the banks. However, the approximation fails during the PSs and, most disappointingly, it is unable to predict whether there is a critical voltage at which the evolution of the superconducting state switches from a quasi-stationary to a quasi-periodic regime. 

By using the adiabatic approximation rather than the full equation (\ref{TDGL}), we are erroneously led to the conclusion that in the case of a nanowire between quasi-normal banks there are no significant PSs for any applied voltage. We should therefore suspect that, analogously, when assuming a priori that stationary KUT is applicable, we may be led to a stationary solution, but this solution can be unstable, even if the KUT procedure converges.

\section*{Acknowledgement}
JB has benefited from extensive correspondence with Alexander Golubov. ES was supported by the Brazilian Agency FAPESP, grant 20/10058-0,
\appendix
\section{Numerical details}\label{appnum}
All the differential equations were solved using the built-in function NDSolve in \textit{Mathematica}, and our strategies were adapted for its use.

In the case of the time-dependent model, we took advantage of the ``method of lines." For this purpose, we added a small term $\delta\,\partial\mu /\partial t$, with $0<\delta\ll 1$, at the right hand side of Eq.\ (\ref{neutral}), and verified that our results were practically independent of $\delta$ and that Eq.\ (\ref{neutral}) was actually obeyed.

In the case of the AA, we found that minimization was more stable than direct use of NDSolve with Dirichlet boundary conditions. For stable branches and fixed $t$, we minimized $F[\psi (0),\partial\psi /\partial x (0),\partial\mu /\partial x (0);t]:=|\psi (L)-re^{iVt/2}|^2/r+|\mu (L)+V/2|^2$; $\psi (L)$ and $\mu (L)$ were evaluated as functions of the values at $x=0$, using NDSolve. Along and near the unstable branches, we fixed $\psi (0)$ rather than $t$ and minimized
$H[\partial\psi /\partial x (0),\partial\mu /\partial x (0);\psi (0)]:=[|\psi (L)|-r]^2/r+|\mu (L)+V/2|^2$; $t$ was then obtained from the phase of $\psi (L)$.

\begin{figure}
\centering
\includegraphics[width=0.9\linewidth]{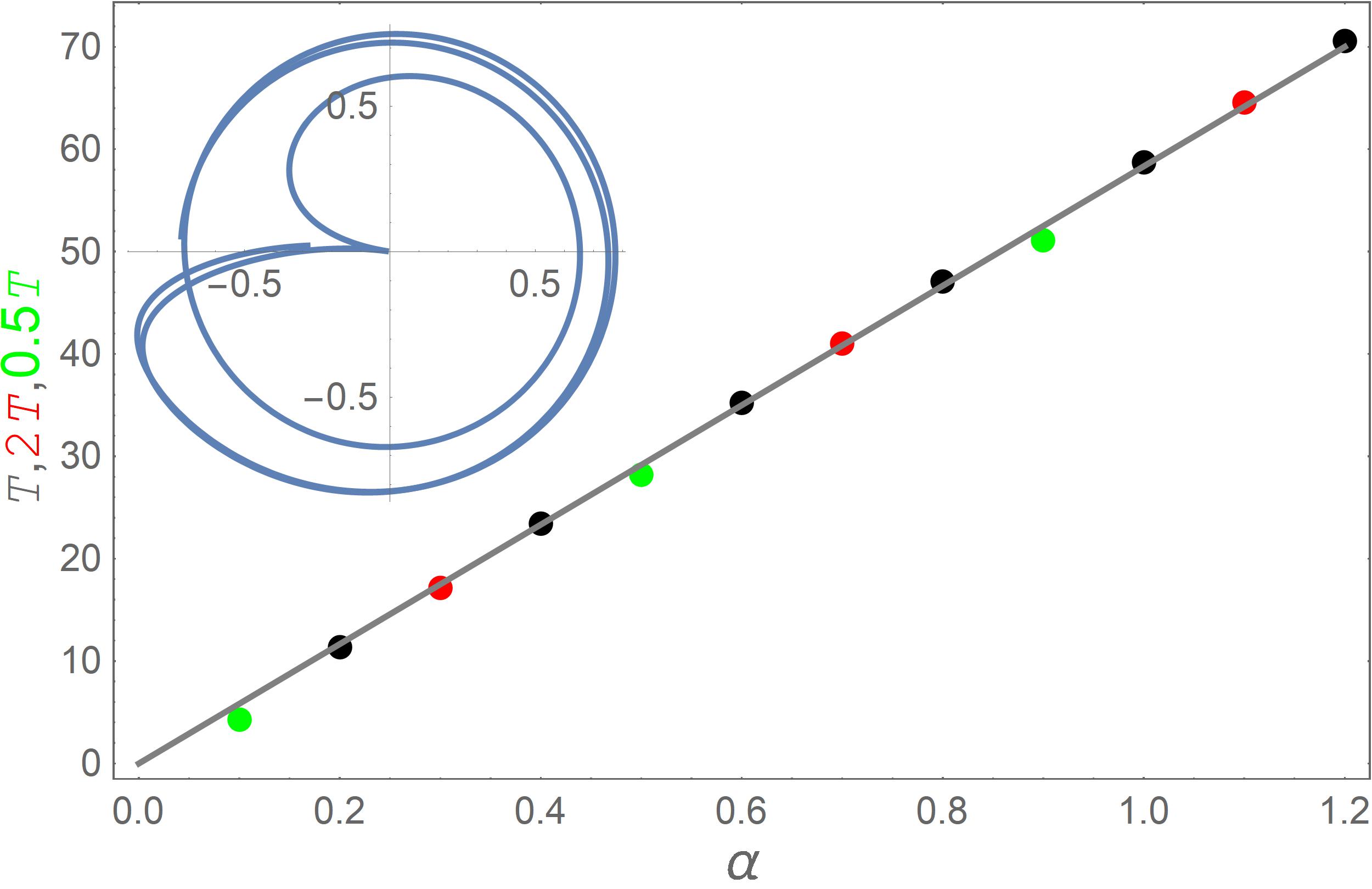}
\caption{Time between consecutive significant phase slips when the time derivative in Eq.\ (\ref{TDGL}) is multiplied by the factor $\alpha$. In all cases, $V_c=2.195$\,. Black: $V=V_c+0.025$\,. Red: $V=V_c+4\times 0.025$; following Eq.\ (\ref{T}), $2T$ is plotted in order to enter a common line. Green: $V=V_c+0.025/4$, $T/2$ is plotted. The straight line describes $T=58.33\alpha$. Inset: polar graph with radius $|\psi (t,0)|$ and angle $Vt/2$ for $V=2.22$ and $\alpha =0.2$.} 
\label{fig:period}
\end{figure} 

\section{Time between phase slips in the adiabatic approximation}\label{timebetween}
We tried to evaluate the time elapsed between consecutive significant PSs using the AA, for applied voltages considerably larger than 2.22, but were unable to find any quasi-periodic regime similar to the one described in Fig.\ \ref{fig:222}. As will become clear below, this approach was doomed to failure.

To overcome this difficulty, we considered ``gradual adiabatization," by multiplying $\partial\psi /\partial t$ in Eq.\ (\ref{TDGL}) by a factor $\alpha$. By doing this, the AA is obtained in the limit $\alpha\rightarrow 0$. In Eq.\ (\ref{T}), $V_c$ and $\beta$ become functions of $\alpha$.

Figure \ref{fig:period} shows that the time between significant PSs can be fitted by taking $V_c$ independent of $\alpha$, and $\beta$ proportional to $\alpha$. This result suggests that periodic significant PSs are not expected in the AA, since, if they were present, the time elapsed between them would have to vanish (whatever that means).

\begin{thebibliography}{00}
\bibitem{Fo}  Foley, C.P. \&  Hilgenkamp, H. Why NanoSQUIDs are important: an introduction to the focus issue. \textit{Supercond. Sci. Technol.} {\bf 22}, 064001 (2009). https://doi.org/10.1088/0953-2048/22/6/064001
\bibitem{We}  Wernsdorfer, W. From micro- to nano-SQUIDs: applications to nanomagnetism. \textit{Supercond. Sci. Technol.} {\bf 22}, 064013 (2009). https://doi.org/10.1088/0953-2048/22/6/064013
\bibitem{optics}You, J. \& Nori, F. Atomic physics and quantum optics using superconducting circuits. \textit{Nature} \textbf{474}, 589–597 (2011). https://doi.org/10.1038/nature10122
\bibitem{Carmine} Granata, C. \&  Vettoliere, A. Nano Superconducting Quantum Interference device: a powerful tool for nanoscale investigations. \textit{Phys. Rep.} {\bf 614}, 1-69 (2016). https://doi.org/10.1016/j.physrep.2015.12.001
\bibitem{LevS}  Levenson-Falk, E. M.,  Antler N. \&  Siddiqi, I. Dispersive nanoSQUID magnetometry. \textit{Supercond. Sci. Technol.} {\bf 29}, 113003 (2016). https://doi.org/10.1088/0953-2048/29/11/113003
\bibitem{KoeRev}  Mart\'{i}nez-P\'{e}rez, M. J. \& Koelle, D. NanoSQUIDs: Basics \& recent advances. \textit{Physical Sciences Reviews} {\bf 2}, 20175001 (2017). https://doi.org/10.1515/psr-2017-5001
\bibitem{PS} Langer, J. S.  \& Ambegaokar, V. Intrinsic resistive transition in narrow superconducting channels. \textit{Phys. Rev.} {\bf 164}, 498-510 (1967). https://doi.org/10.1103/PhysRev.164.498
\bibitem{science} Madan I,\textit{et al.}, Nonequilibrium optical control of dynamical states in superconducting nanowire circuits. \textit{Science Adv} \textbf{4} eaao0043 (2018).  https://www.science.org/doi/10.1126/sciadv.aao0043
\bibitem{Jackel} Dolan, G. J.  \& Jackel, L. D. Voltage measurements within the nonequilibrium region near phase-slip centers. \textit{Phys. Rev. Lett.} \textbf{39}, 1628 (1977). https://doi.org/10.1103/PhysRevLett.39.1628
\bibitem{Q} Mooij, J.E. \& Hartmans C.J.P.M. Phase-slip flux qubits. \textit{New J. Phys } \textbf{7} 219 (2005).
\bibitem{IEEE}  Kitaygorsky, J.  \textit{et al.} Origin of dark counts in nanostructured NbN single-photon detectors. \textit{ IEEE Transac. Appl. Supercond.} \textbf{15} 545-548 (2005). 
\bibitem{stat} Berger J. Stationary nano-SQUID: theoretical investigation and feasibility analysis. \textit{J. Phys.: Condens. Matter} \textbf{29} 29LT01 (2017). https://doi.org/10.1088/1361-648X/aa75c4
\bibitem{Koby} Rubinstein, J., Sternberg, P. \& Ma, Q.  Bifurcation diagram and pattern formation of phase slip centers in superconducting wires driven with electric currents. \textit{Phys. Rev. Lett.} {\bf 99}, 167003 (2007). https://doi.org/10.1103/PhysRevLett.99.167003
\bibitem{lossless} L. Kramer and A. Baratoff, Lossless and dissipative current-carrying states in quasi-one-dimensional superconductors. \textit{Phys. Rev. Lett.} \textbf{38}, 518 (1977). https://doi.org/10.1103/PhysRevLett.38.518 
\bibitem{thermal} McCumber, D. E. \& Halperin, B. I.  Time scale of intrinsic resistive fluctuations in thin superconducting wires. \textit{Phys. Rev. B} {\bf 1}, 1054-1070 (1970). https://doi.org/10.1103/PhysRevB.1.1054
\bibitem{q1}Lau, C. N.,  Markovic, N., Bockrath, M., Bezryadin, A. \& Tinkham, M.  Quantum phase slips in superconducting nanowires. \textit{Phys. Rev. Lett.} {\bf 87}, 217003 (2001). https://doi.org/10.1103/PhysRevLett.87.217003 
\bibitem{q2}  Arutyunov, K.Yu., Golubev, D. S. \&  Zaikin, A. D. Superconductivity in one dimension. \textit{Physics Reports} {\bf 464} 1-70  (2008). https://doi.org/10.1016/j.physrep.2008.04.009
\bibitem{KW1} Kramer, L. \&  Watts-Tobin, R. J. Theory of dissipative current-carrying states in superconducting filaments. \textit{Phys. Rev. Lett.} {\bf 40}, 1041-1044 (1978). https://doi.org/10.1103/PhysRevLett.40.1041
\bibitem{KW2}Watts-Tobin, R.J., Krähenbühl, Y. \& Kramer, L. Nonequilibrium theory of dirty, current-carrying superconductors: phase-slip oscillators in narrow filaments near $T_c$. \textit{J. Low Temp. Phys.} {\bf 42}, 459–501 (1981). https://doi.org/10.1007/BF00117427
\bibitem{Ivlev} Ivlev, B. I. \& Kopnin, N. B. Electric currents and resistive states in thin superconductors, \textit{Adv. Phys.} {\bf 33}, 47-114 (1984). https://doi.org/10.1080/00018738400101641
\bibitem{Rangel}Kramer, L. \& Rangel, R. Structure and properties of the dissipative phase-slip state in narrow superconducting filaments with and without inhomogeneities. \textit{J. Low Temp. Phys.} {\bf 57}, 391–414 (1984). https://doi.org/10.1007/BF00681200
\bibitem{me} Berger, J. Influence of the boundary conditions on the current flow pattern along a superconducting wire. \textit{Phys. Rev. B} {\bf 92}, 064513 (2015). https://doi.org/10.1103/PhysRevB.92.064513
\bibitem{physc} Kim, J.,  Rubinstein, J., \&  Sternberg, P. Length/voltage phase diagram for a thin superconducting wire subjected to an applied voltage. \textit{Physica C} {\bf 470}, 630–634 (2010). https://doi.org/10.1016/j.physc.2010.06.004
\bibitem{Micho1} Vodolazov, D.Y., Peeters, F.M., Piraux, L.,  M\'{a}t\'{e}fi-Tempfli, S. \& Michotte, S.  Current-voltage characteristics of quasi-one-dimensional superconductors:
an S-shaped curve in the constant voltage regime. \textit{Phys. Rev. Lett.} {\bf 91}, 157001 (2003). https://doi.org/10.1103/PhysRevLett.91.157001
\bibitem{Micho2} Michotte, S., M\'{a}t\'{e}fi-Tempfli, S., Piraux, L., Vodolazov, D. Y. \& Peeters, F. M. Condition for the occurrence of phase slip centers in superconducting nanowires under applied
current or voltage. \textit{Phys. Rev. B} {\bf 69}, 094512 (2004). https://doi.org/10.1103/PhysRevB.69.094512
\bibitem{Belzig} Belzig, W.,  Wilhelm, F. K., Bruder, C., Schön, G. \& Zaikin, A. D. Quasiclassical Green’s function approach to mesoscopic superconductivity. \textit{Superlatt. Microstruct.} {\bf 25}, 1251-1288 (1999). https://doi.org/10.1006/spmi.1999.0710. Corrigendum: \textit{Superlatt. Microstruct.} {\bf 35}, 157 (2004) https://doi.org/10.1016/j.spmi.2004.01.003
\bibitem{Kamenev} Kamenev, A. \& Levchenko, A.  Keldysh technique and non-linear $\sigma$-
model: basic principles and applications, \textit{Adv. Phys.} {\bf 58}, 197-319 (2009). http://dx.doi.org/10.1080/00018730902850504
\bibitem{SQUIPT} Giazotto, F., Peltonen, J., Meschke, M. et al. Superconducting quantum interference proximity transistor. \textit{Nature Phys} \textbf{6}, 254–259 (2010). https://doi.org/10.1038/nphys1537
\bibitem{Brink} A. Brinkman and A. A. Golubov, Crossed Andreev reflection in diffusive contacts: Quasiclassical Keldysh-Usadel formalism. \textit{Phys. Rev. B} \textbf{74}, 214512 (2006). https://doi.org/10.1103/PhysRevB.74.214512
\bibitem{cph} A. A. Golubov, M. Yu. Kupriyanov, and E. Il’ichev  The current-phase relation in Josephson junctions. \textit{Rev. Mod. Phys.} \textbf{76}, 411 (2004). 
\bibitem{Buzdin} Buzdin, A.I. Proximity effects in superconductor-ferromagnet heterostructures. \textit{Rev. Mod. Phys.} \textbf{77}, 935 (2005). https://doi.org/10.1103/RevModPhys.77.935

\bibitem{Boo} Boogaard, G. R., Verbruggen, A. H., Belzig, W. \&  Klapwijk, T. M. Resistance of superconducting nanowires connected to normal-metal leads. \textit{Phys. Rev. B} {\bf 69}, 220503 (2004). https://doi.org/10.1103/PhysRevB.69.220503
\bibitem{Keizer} Keizer, R. S., Flokstra, M. G., Aarts, J. \& Klapwijk, T. M. Critical voltage of a mesoscopic superconductor. \textit{Phys. Rev. Lett.} {\bf 96}, 147002 (2006). https://doi.org/10.1103/PhysRevLett.96.147002
\bibitem{Pekola} Vercruyssen, N., Verhagen, T. G. A., Flokstra, M. G., Pekola, J. P. \& Klapwijk, T. M. Evanescent states and nonequilibrium in driven superconducting nanowires. 
\textit{Phys. Rev. B} {\bf 85}, 224503 (2012). https://doi.org/10.1103/PhysRevB.85.224503
\bibitem{Andreev} Dolgirev, P.E., Kalenkov, M.S. \& Zaikin, A.D. Interplay between Josephson and Aharonov-Bohm effects in Andreev interferometers. \textit{Sci Rep} \textbf{9}, 1301 (2019). https://doi.org/10.1038/s41598-018-37653-w
\bibitem{Golub} Brinkman, A., Golubov, A. A., Rogalla, H., Wilhelm, F. K. \&  Kupriyanov, M. Yu. Microscopic nonequilibrium theory of double-barrier Josephson junctions.
\textit{Phys. Rev. B} {\bf 68}, 224513 (2003). https://doi.org/10.1103/PhysRevB.68.224513
\bibitem{tunnel} Bezuglyi, E. V., Vasenko, A. S., Shumeiko, V. S. \& Wendin, G. Nonequilibrium effects in tunnel Josephson junctions.
\textit{Phys. Rev. B} {\bf 72}, 014501 (2005). https://doi.org/10.1103/PhysRevB.72.014501

\end{thebibliography}


\end{document}